\documentclass[pra,reprint,aps,showpacs,10pt,floatfix]{revtex4-1}

\usepackage{amsfonts}
\usepackage{amsmath}
\usepackage{graphicx}
\usepackage[latin1]{inputenc}
\usepackage{color}
\usepackage{bbold}
\usepackage{braket}
\usepackage{mathtools}

\newcommand{\be}{\begin{equation}}
\newcommand{\ee}{\end{equation}}
\newcommand{\BE}{\begin{equation*}}
\newcommand{\EE}{\end{equation*}}
\newcommand{\bea}{\begin{eqnarray}}
\newcommand{\eea}{\end{eqnarray}}
\newcommand{\BEA}{\begin{eqnarray*}}
\newcommand{\EEA}{\end{eqnarray*}}

\DeclareMathOperator{\tr}{Tr}
\begin{document}
\title{What can we learn from the dynamics of entanglement and quantum discord in the Tavis--Cummings model?}

\author{Juliana Restrepo}
\affiliation{Sistemas Complejos, Universidad Antonio Nari\~{n}o, Medell\'{i}n, Colombia}

\author{Boris A. Rodr\'{i}guez}
\affiliation{Instituto de F\'{i}sica, Universidad de Antioquia UdeA, Calle 70 No. 52-21, Medell\'{i}n, Colombia}

\begin{abstract}
We revisit the problem of the dynamics of quantum correlations in the exact Tavis--Cummings model. We show that many of the dynamical features of quantum discord attributed to dissipation are already present in the exact framework and are due to the well known non-linearities in the model and to the choice of initial conditions. Through a comprehensive analysis, supported by explicit analytical calculations, we find that the dynamics of entanglement and quantum discord are far from being trivial or intuitive. In this context, we find states that are indistinguishable from the point of view of entanglement and distinguishable from the point of view of quantum discord, states where the two quantifiers give opposite information and states where they give roughly the same information about correlations at a certain time. Depending on the initial conditions, this model exhibits a fascinating range of phenomena that can be used for experimental purposes such as:  Robust states against change of manifold or dissipation, tunable entanglement states and states with a counterintuitive sudden birth as the number of photons increase. We furthermore propose an experiment called quantum discord gates where discord is zero or non-zero depending on the number of photons. 
\end{abstract}

\date{\today} 
\maketitle

\section{Introduction}

The quantum correlations arising from the superposition principle have been the source of a long-standing and heated debate since the birth of quantum mechanics \cite{EPR, Bohr, Olival}. Indeed, when Schr\"odinger declared in 1935 that \textquotedblleft entanglement is the characteristic trait of quantum mechanics\textquotedblright \cite{Schrodinger} it was believed that the key quantum correlation was entanglement \cite{HorodeckiRev,EntanglementRev} and many efforts were devoted to quantify entangled vs separable states \cite{Werner}. This main role of entanglement was encouraged by the Bell theorem \cite{Bell}, its experimental verifications \cite{Aspect, Zeilinger, Gisin} and its use as a main resource in quantum information \cite{HorodeckiRev} and quantum computation \cite{Nielsen} tasks. However, entanglement is not the only quantum correlation encoded in a quantum state and it was found recently, that certain tasks, e.g. \textit{nonlocality without entanglement} \cite{Bennet, Bartlett} and \textit{quantum speedup with separable states} \cite{Braunstein, Lanyon} can be done without using entanglement as a resource. %OK, verificar italica y comillas

From this point of view, it is desirable to investigate other quantifiers of quantum correlations. Recently, Olivier and Zurek introduced the quantum discord \cite{Zurek, Vedral} as the difference between two possible quantum extensions of the classical mutual information. This new quantity has been broadly studied in the context of quantum information \cite{Winter,  Datta, Dakic, Piani}, quantum cryptography \cite{Pirandola} and quantum metrology \cite{Adesso2014}.  Even though the quantum and classical information encoded in quantum discord cannot be directly compared \cite{Modi}, several theoretical efforts have been made in order to understand how they are contained in a quantum state \cite{Acin, Modi, Modi2012}, and it is presently understood that the quantum discord quantifies in a more subtle way the quantum correlations in mixed states through the state disturbance induced by local measurements \cite{Streltsov, Piani2012}.
% Verificar redaccion de la frase "even though" y verificar que sea cierta, citar lo de comparar, verificar italica!

The dynamics of quantum entanglement have been exhaustively studied \cite{Buchleitner}. One of the important features of entanglement is the ``entanglement sudden death" (ESD) phenomenon \cite{YuEberly, Davidovich, Kimble}. This process describes the disentaglement of a pair of qubits exposed to an environment in a finite time, and it depends strongly on the initial state of the system and its interaction with the environment. On the other hand, the dynamics of quantum discord have been mainly studied under decoherence and dissipation scenarios. For a pair of qubits coupled to a Markovian environment \cite{Fanchini2009, Celeri, Serra, Mazzola}, the quantum discord was sometimes seen to have an exponential decay and an asymptotic vanishing \cite{Fanchini2009, Acin, Mazzola}. The effect of non-Markovian environment has also been studied in \cite{Fanchini2010}. The ensuing quantum discord  was found to vanish only at discrete instants of time. In this context it is relevant to ask up to which extend these dynamical phenomena are due to the dissipative dynamics or to the non-linearity contained in the matter--field dipole--type interaction in the models considered. 

In this paper, we address the dynamics of quantum discord and entanglement in the Tavis--Cummings exact Hamiltonian. We show that some of the dynamical features observed for quantum discord considering dissipation are already present in the exact model and do not need assumption on the coupling or the nature of the environment. Though a direct comparison in amount of discord and entanglement in meaningless, because the measure of discord and entanglement does not coincide, we make some general remarks on the different correlations embedded in a quantum state and the implications on the geometry of Hilbert space following the interpretation of K. Modi et al., \cite{Modi}. Furthermore we find surprising effects for certain initial conditions such as the possibility of building quantum discord gates (zero or non zero discord depending on the number of photons), robust states against dissipation and states whose entanglement counterintuitively augments with the number of photons, among others.
%We find ***cosas bacanisimas ***  and Furthermore we propose? how nuevos, interesantes results can be used for ** for quantum *** information experiments that use quantum discord and or entanglement to ***. \\ % En las referencias esta redactado regulimbis y falta conector. Despus falta contundencia en la primera frase In this context. Tambien falta poner mejor lo del final de los experimentos y decidir que resultados ponemos / si ponemos (esto depende de la revista, no?) aqui ya. Hamilton o hamilton. Que tanto hablar del nombre del modelo aqui? queda como ambiguo

The paper is organized as follows: we present the exact model in Sec II  followed by the definition and a discussion of the entanglement and quantum discord in Sec. III. We then present our results for the dynamics of correlations of different initial conditions in Section IV. Finally, Section V summarizes the results and draws conclusions.% Muy cortico, falta poner que al final discutimos los resultados desde el punto de visa de no se que mierda

\section{Tavis--Cummings model}
We present the model Hamiltonian used to study the dynamics of entanglement and quantum discord. The Hamiltonian that describes the interaction between two non-interacting two--level systems (2-TLS) and a single--mode cavity field, is the so-called Tavis--Cumings Hamiltonian (TC) \cite{TC}, a generalization of the Jaynes-Cummigs Hamiltonian (JC) \cite{JC, ReviewJC}. The JC and TC models of matter-light interaction are the theoretical cornerstone in an variety of quantum related areas such as quantum optics \cite{Walls-Milburn, Special}, cavity QED (CQED) \cite{Haroche, Walther}, trapped ions \cite{Wineland1996, WinelandRMP}, quantum information and quantum computation \cite{Nico, Azuma, Molmer}, circuit QED \cite{Schoelkopf, Schoelkopf2004, Mooij, Takayanagi}, semiconductor quantum optics \cite{KiraKoch, Forchel, Deppe, Elena, Elena2012, us} and cavity opto-mechanics \cite{Marquardt, Milburn-Woolley}. In the dipole and rotating wave approximation, the TC Hamiltonian can be written as: 
\be
H =  \frac{\omega}{2}(\sigma^{z}_1+\sigma^{z}_2)+ \omega_{0} a^\dagger a +g \sum_{i=1}^2  \left( a^\dagger \sigma_{i}^{-}+ a \sigma_{i}^{\dagger}\right),
\label{TC-h}
\ee
where the first two terms are the 2-TLS and photon energies respectively, $a$ ($a^\dagger$) are the creation and annihilation operators of the single-mode cavity field, and $\sigma_i^\dagger=\ket{+}_i\prescript{}{i}{\bra{-}}$ and $\sigma_i^-=\ket{-}_i\prescript{}{i}{\bra{+}}$ are the TLS pseudo--spin flip operators that connect the ground $\ket{-}_i$ an excited $\ket{+}_i$ states of the i-th TLS with energies  $- \omega/2$ and $\omega/2$ respectively. The interaction Hamitonian describes the dipole interaction between the 2-TLS and the field, $g$ is the light-matter coupling constant. We take $\hbar=1$ and the resonance condition $\omega-\omega_0=0$.
% FALTA, revisar italica pseudo, ket bra falta una i, omega y   omega_0

The Tavis--Cummings Hamitonian has a conserved quantum number, the so-called excitation manifold number, given by $\Lambda = N + N_e$, where $N=a^\dagger a$ is the number of photons and $N_e=\sum_{i=1}^2 \sigma_i^{\dagger} \sigma_i^{-}$ is the number of TLS in the excited state. Using this symmetry, we can separate the evolution of the total density operator $\rho^{T}(t)$ in disjoint excitation manifolds $\Lambda_{n}\doteq \{ \ket{++}_{n-1}, \ket{+-}_{n}, \ket{-+}_{n}, \ket{--}_{n+1}\}$ and calculate the exact solution for total density operator of the system:
\be
%\rho^{\mathrm{2-TLS-Field}}(t) \doteq \rho^{S}(t)=\hat{U}(t) \rho^{S}(0) \hat{U}^\dagger(t)
\rho^{T}(t) =\hat{U}(t) \rho^{T}(0) \hat{U}^\dagger(t),
\ee
where $\rho^{T}(t)=\oplus_{\substack{n}} \rho^T(t)|_{\Lambda_{n}}$ and  $\hat{U}(t)=\oplus_{\substack{n}} \hat{U}(t)|_{\Lambda_{n}} $. The operators $\rho^T(t)|_{\Lambda_{n}}$ and $ \hat{U}(t)|_{\Lambda_{n}}$ are the density and the time evolution operators in manifold $\Lambda_n$ respectively. The latter can be written as \cite{Kim2002,Puri}:
\begin{widetext}
\be
\hat{U}(t)|_{\Lambda_{n}}= e^{-i \omega n t} \times
\begin{pmatrix}
1+\frac{n C_1(t)}{2n+1}  & - \frac{ i  \sqrt{n} C_2(t)} {\sqrt{2(2n+1)}} &  - \frac{ i  \sqrt{n} C_2(t)} {\sqrt{2(2n+1)}}  & \frac{\sqrt{n(n+1)} C_1(t)}{2n+1} \\
- \frac{ i  \sqrt{n} C_2(t)} {\sqrt{2(2n+1)}} & 1 + \frac{C_1(t)}{2} & \frac{C_1(t)}{2} & - \frac{ i  \sqrt{n+1} C_2(t)} {\sqrt{2(2n+1)}} \\
- \frac{ i  \sqrt{n} C_2(t)} {\sqrt{2(2n+1)}} & \frac{C_1(t)}{2} & 1 + \frac{C_1(t)}{2} & - \frac{ i  \sqrt{n+1} C_2(t)} {\sqrt{2(2n+1)}} \\
\frac{\sqrt{n(n+1)} C_1(t)}{2n+1} & - \frac{ i  \sqrt{n+1} C_2(t)} {\sqrt{2(2n+1)}} & - \frac{ i  \sqrt{n+1} C_2(t)} {\sqrt{2(2n+1)}} & 1+\frac{(n+1) C_1(t)}{2n+1}
\end{pmatrix}.
\label{exact}
\ee
\end{widetext}

\noindent
where the time dependent functions $C_1(t)=\cos(2\pi\Omega_{R} t)-1$ and $C_2(t)=\sin(2\pi\Omega_R t)$ oscillate with an effective Rabi frequency $\Omega_{R}=g\sqrt{4n+2}$. The above expression enables us to calculate the dynamics for any initial state of the 2-TLS + light system, pure or mixed. Restricting  ourselves to initial conditions $\rho^{T}(0)$ in the excitation manifold $\Lambda_n$, we can write the total 2-TLS + light density operator in the $\Lambda_n$ manifold. This restriction results in a global phase in Eq. (\ref{exact}) and the subsequent evolution will not depend on $\omega$. In this paper we focus on the 2-TLS which are formally a two-qubit system. Taking the partial trace on the field states $\rho(t)= \tr_{\mathrm{Field}}{\rho^T(t)}$ and using the usual basis for the 2-TLS Hilbert space $\mathcal{H} \doteq \{ \ket{1} = \ket{++}, \ket{2} = \ket{+-}, \ket{3} = \ket{-+}, \ket{4} = \ket{--}\}$, we obtain the reduced two-qubit density matrix:
\bea
\rho(t)=
\begin{pmatrix}
\rho_{11}(t) & 0 & 0 & 0 \\
0 & \rho_{22}(t)& \rho_{23}(t) & 0 \\
0 & \rho_{32}(t) & \rho_{33}(t) & 0 \\
0 & 0 & 0 & \rho_{44}(t)
\end{pmatrix}.
\label{pmatrixx}
\eea
The above matrix has an X structure \cite{Eberly2007} which will turn out to be very useful to compute the dynamics of quantum discord. % FALTA no se si se debe volver a definir la base... queda obvio o no???? me parece que toda la notacion igual confunde un poco y faltan cosas entre lineas, como terminar???

\section{Classical and quantum correlations}
\subsection{Entanglement}
Entanglement is a measure of the non-separability of the quantum state of a composite system and it is, in general, a difficult quantity to compute \cite{HorodeckiRev, EntanglementRev}. As a resource in quantum information and computation tasks, the entanglement expresses the maximum number of Bell pairs that it is possible to obtain from the quantum state to be used for quantum tasks. From the work of Wootters \cite{Wootters} in the nineties, it is well known that the entanglement for a pair of qubits can be quantified in the concurrence, a function which has a closed simple form for any state of the TLS given the density matrix $\rho$ that describes them.

For a two-qubit system the concurrence is given by \cite{Wootters} $\max\,\{0,\Lambda(t)\}$, where $\Lambda(t)=\lambda_1(t)-\lambda_2(t)- \lambda_3(t)-\lambda_4(t)$ and $\lambda_i(t)$ are the square roots, ordered in decreasing value, of the eigenvalues of the matrix $\rho(t)(\sigma_2 \otimes \sigma_2) \rho^{\ast}(t) (\sigma_2 \otimes \sigma_2)$. $\rho^{\ast}(t)$ is the complex conjugate of the two-qubit density matrix $\rho(t)$ and $\sigma_2$ the second Pauli matrix. As the reduced two-qubit density matrix we obtain to evaluate concurrence has an X structure with $\rho_{14}(t)=0$, cf. Eq.~(\ref{pmatrixx}), the concurrence has a simple analytic expression $C(t)= 2 \max\,\{0, |\rho_{23}(t)|\sqrt{\rho_{11}(t)\rho_{44}(t)}\}$.

\subsection{Quantum discord}

Following Olivier and Zurek \cite{Zurek}, we can quantify the total amount of classical and quantum correlations present in a bipartite quantum system $\rho^{AB}$ by means of the quantum mutual information, an information-theoretic measure of the total correlation in a bipartite quantum state:
\be
\mathcal{I}(\rho^{AB}) = S(\rho^{A}) + S(\rho^{B}) - S(\rho^{AB}),
\ee
where $S(\rho) = - \tr(\rho \log \rho)$ is the von-Neumann entropy. The quantum mutual information may be written as a sum of classical correlations $\mathcal{C}(\rho^{AB})$ and the quantum discord $D(\rho^{AB})$:
\be
D(\rho^{AB})=\mathcal{I}(\rho^{AB}) - \mathcal{C}(\rho^{AB}),
\label{resta}
\ee
with the former given by \cite{Zurek, Vedral}:
\bea
\mathcal{C}(\rho^{AB}) &=& \underset{\{ \Pi_k^B\}}{\max} \left( \mathcal{I}(\rho^{AB}|\Pi_k^B) \right); \nonumber \\
  &=& S(\rho^A)-\underset{\{ \Pi_k^B\}}{\min} \left( S(\rho^{AB}| \{\Pi_k^B\}) \right),
 \label{clasiquito}
\eea
where $ \mathcal{I}(\rho^{AB}|\Pi_k^B)$ is the  quantum conditional mutual information of a measurement, with $\{ \Pi_k^B\}$ a complete set of projection operators, in the subsystem $B$, and $\rho^{AB}| \{\Pi_k^B\} = \tr_B(\Pi_k^B \rho^{AB} \Pi_k^B)/p_k$ is the residual state of the subsystem $A$ after the measurement with result $k$ and probability $p_k = \tr_{AB}(\Pi_k^B \rho^{AB} \Pi_k^B)$. Due to the complex optimization procedure involved in the definition of classical correlations, the quantum discord is usually intractable to compute for a general state. However recently efforts in the two-qubit case has shown that it is possible to obtain a closed expression for the quantum discord of a general two-qubit state \cite{Adesso, Zhou}.  
%Despite a small bounded error \cite{Huang, Namkug}, 
For the class of two-qubit X-states it is possible to obtain an analytical expression \cite{Li, Ali,Luo} for the quantum discord:
\be
D(\rho^{AB})= \min \left( D_{\sigma_x^B}(\rho^{AB}), D_{\sigma_z^B}(\rho^{AB}) \right),
\label{discord}
\ee
where $D_{\Pi_k^B}(\rho^{AB}) = \mathcal{I}(\rho^{AB}) - \mathcal{I}(\rho^{AB}|\Pi_k^B)$.  Recently, Chen and Huang \cite{Chen,Huang} introduced a theorem that states that expression (\ref{discord}) is exact if: 
\begin{equation}
(|\rho_{23}|+|\rho_{14}|)^2\leq (\rho_{11}-\rho_{22})(\rho_{44}-\rho_{33});
\label{uno}
\end{equation}
or
\begin{equation}
|\sqrt{\rho_{11}\rho_{44}}-\sqrt{\rho_{22}\rho_{33}}|\leq |\rho_{23}|+|\rho_{14}|,
\label{dos}
\end{equation}
with $\rho_{ij}=\rho_{ij}(t)$, $i,j=1,\ldots, 4$ defined in Eq. (\ref{pmatrixx}). In the first case the minimum is $D_{\sigma_z^B}(\rho^{AB})$ and in the latter case the minimum is $D_{\sigma_x^B}(\rho^{AB})$. There exist the possibility of not satisfying eqs (\ref{uno}) and (\ref{dos}), in this case Huang \cite{Huang} bounded the error for the quantum discord to $0.0021$. For most of the initial conditions and range of parameters considered in this paper there are small regions in time, of order $0.02t_R$, where the quantum discord is neither $D_{\sigma_x^B}(\rho^{AB})$ nor $D_{\sigma_x^B}(\rho^{AB})$ but the effect of considering the other minimum is negligible given the order of the correlations measured by discord. 
%***However, considering the other minimization would only imply small changes in the dynamics***.  
There are two important results in our paper that are of the same order of magnitude of the error, in both cases we specifically comment on the validity of expression (\ref{discord}) for quantum discord.
% but the changes would be negligeable? con respecto a la dinamica . Hay dos casos que es importante mencionar: para algunos parametros especiales pueden haber diferencias en tiempos considerables ($0.1t_R$) y dos: a veces podrian haber en vez de una discontinuidad una doble discontinuidad (pero hasta aca no he hablado de discontinuidades). %habria sido bacano decir que por ejemplo: en 14+ siempre es chiquito el error, en 23+, 14-, ali2a hay regiones muy grandes de error y unas son cerca de la discontinuidades si es que las hay, para 23+ para alfa 0 y alfa 1 siempre es sigma x la que mide, para 23- y cualquier parametro siempre es sigma x la que mide, en las regiones de parametros importantes para el experimento de discord gates siempre es sigma z la que aparece***.

\section{Dynamics of quantum correlations}
We now use the definitions presented in the previous section to calculate the concurrence $C(t)$ and the quantum discord $D(t)$ in the Tavis--Cummings model cf. Eq. (\ref{TC-h}). In order to comprehend the dynamics of quantum correlations we consider various families of pure and mixed initial conditions. For each of them, we compare the two quantities and discuss how the results might be used for quantum computation purposes.
%%%%%%%%%%%%%%%%%%%%%%%%%%%%%%%%%%%%%%%%%%%%%%%%	%%%%
%%%%%%%%%%%%%%%%%%%%%%%%%%%%%%%%%%%%%%%%%%%%%%%%%%%%
%%%%%%%%%%%%%%%%%%%%%%%%%%%%%%%%%%%%%%%%%%%%%%%%%%%%
%%%%%%%%%%%%%%%%%%%%%%%%%%%%%%%%%%%%%%%%%%%%%%%%%%%%
%%%%%%%%%%%%%%%%%%%%%%%%%%%%%%%%%%%%%%%%%%%%%%%%%%%%
%%%%%%%%%%%%%%%%%%%%%%%%%%%%%%%%%%%%%%%%%%%%%%%%%%%%
% questions that remain
% deberia mencionar como s mueven los vanishing times y los kink times del discord? La tengo clara (ver tavis_v4.nb) pero en ninguna parte lo menciono
% deberia quitar lo que "no se puede comparar de Modi" del principio del ultimo parrafo?
% deberia mostrar la robustez del discord y la concurrencia en una grafica de D/C con n=1 y n=100 fotones?

\subsection{Family of Bell states}
%******
%figure
\begin{figure}
\includegraphics[width=0.45\textwidth]{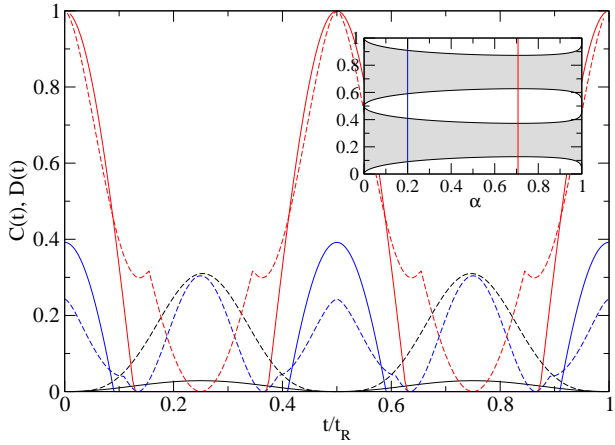}
\caption{\label{0313fig23+}(color online) Concurrence (bold lines) and quantum discord (dashed lines) as a function of time for initial condition $\ket{\psi^+_\alpha}_1$, with $\alpha=0$ (black), $\alpha=0.2$ (blue) and $\alpha=1/\sqrt{2}$ (red). The inset shows the dependence of collapse and revival times $t_i/t_R$, $i=1,2,3,4$ as a function of $\alpha$ for manifold $\Lambda_1$. The shaded areas in the inset correspond to the ESD and the vertical lines correspond to the values of $\alpha$ considered in the large figure.}
\end{figure}
%figure
%*****
As a first example we take the partially entangled pure states:
\begin{equation}
%\ket{\psi^\pm_\alpha}=\alpha \ket{2}_n\pm\sqrt{1-\alpha^2}\ket{3}_n
\ket{\psi^\pm_\alpha}_n=\alpha \ket{+-}_n\pm\sqrt{1-\alpha^2}\ket{-+}_n,
\end{equation}
with $0\leq\alpha\leq1$. For $\alpha=1/\sqrt{2}$ these correspond to the usual Bell States which are maximally entangled. For such states quantum discord and any measurement of entanglement coincide \cite{Ali}.

\begin{figure}
\includegraphics[width=0.45\textwidth]{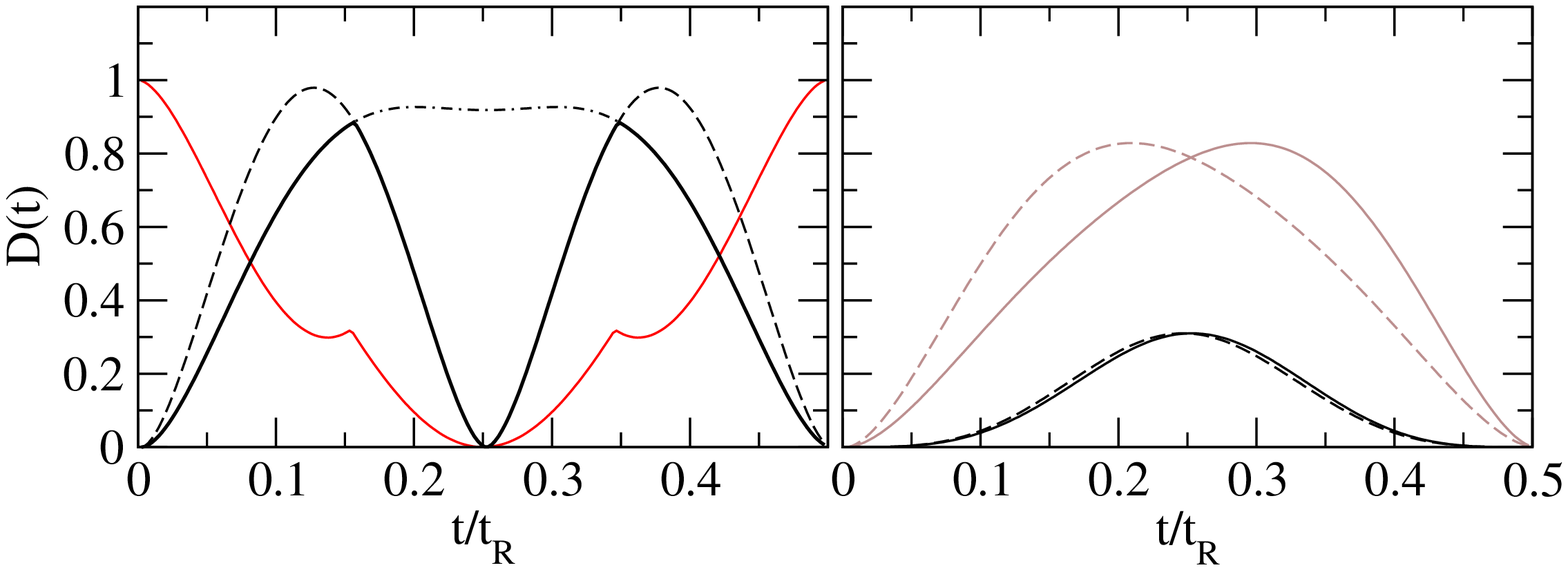}
\caption{\label{0415disc}(color online) Left panel: $S(\rho_{AB}|\sigma_z^B)$ (black-dashed), $S(\rho_{AB}|\sigma_x^B)$ (black-point dashed),  the minimum ${\min} \left( S(\rho^{AB}| \{\sigma_z^B,\sigma_x^B\}) \right)$ (black) and quantum discord (red)  as a function of time for initial condition $\ket{\psi^+_{1/\sqrt{2}}}_1$. Right panel: ${\min} \left( S(\rho^{AB}| \{\sigma_z^B,\sigma_x^B\}) \right)=S(\rho_{AB}|\sigma_x^B)$ (brown) and quantum discord $D(t)$ (black) as a function of time for initial conditions $\ket{\psi^+_0}_1$ (bold) and $\ket{\psi^+_1}_1$ (dashed). The vertical scale on both panels is the same.}
\end{figure}

%*****
We first focus on the results for the family of initial conditions $\ket{\psi^+_\alpha}_n$. For $0<\alpha<1$ (cf Fig. \ref{0313fig23+}) the concurrence $C(t)$ starts at a maximum $C(0)$ given by:
\begin{equation}
C(0)\equiv g(\alpha)=2\alpha\sqrt{1-\alpha^2}.
\label{galpha}
\end{equation} 
An entanglement sudden death (ESD) appears at $0<t_1<t_R/4$ and a sudden revival at  $t_R/4<t_2<t_R/2$  augmenting to the same value $g(\alpha)$ at $t=t_R/2$. The collapse ($t_1$ and $t_3$) and revival ($t_2$ and $t_4$) times as a function of $\alpha$ in the interval $\left[0,t_R\right]$ are plotted in the inset of Figure \ref{0313fig23+}. The discord, on the other hand, only vanishes at discrete times. For $0<\alpha<1$, it starts at a maximum:
%$\alpha\neq\tfrac{1}{\sqrt{2}}$
\begin{equation}
D(0)=\left(\alpha ^2-1\right) \log_2 \left(1-\alpha ^2\right)-\alpha ^2 \log_2 \left(\alpha ^2\right).
\label{D0}
\end{equation}
As time increases, it vanishes and oscillates to a second maximum at $t=t_R/4$ symmetrically vanishing again and oscillating to the maximum at $t_R/2$. Depending on the number of initial photons and the initial condition the quantum discord at $t_R/4$ can be larger, equal or smaller than the quantum discord at $t_R/2$. For example, for the manifold $\Lambda_1$, for $\alpha\simeq0.22$ and $\alpha\simeq0.97$ the discord at $t_R/4$ is equal to the discord at $t_R/2$ and for $\alpha\simeq0.18$ one obtains the maximum quantum discord at $t_R/4$. As can be seen in Figure  \ref{0313fig23+} the discord has discontinuities in the derivative before vanishing at $t<t_R/4$ and after vanishing at $t>t_R/4$. These discontinuities were already reported by Maziero et al., when they considered an open system obeying the Tavis--Cummings Hamiltonian coupled to non-Markovian baths \cite{Serra}. In general, one can say that they are a common feature in the dynamics of quantum discord and are due to the minimization of entropies in expression (\ref{discord}) (see left panel in Fig. \ref{0415disc}). Furthermore, in the Tavis--Cummings model they are already present in the exact dynamics.
%*****
%figure
%figure
%*****

The dependence of dynamics on the manifold $\Lambda_n$, apart from the trivial dependency of the Rabi frequency, is only in the elements $\rho_{11}(t)$ and $\rho_{44}(t)$ of the reduced density matrix  and is of order $\mathcal{O}(1/n)$. Both for the concurrence and the quantum discord this leads to corrections of order $\mathcal{O}(1/n^2)$ or $\mathcal{O}(1/n^4)$. In this way, we see that for the family under consideration 
% discord this leads to classical and quantum correlations with corrections of order $\mathcal{O}(1/n^2)$ or $\mathcal{O}(1/n^4)$ for some times and for the concurrence of order $\mathcal{O}(1/n^2)$ for all times ***deje solo el 2***. This implies that for this initial condition
the correlations measured by discord and concurrence are robust when changing the manifold and survive as $n\to\infty$. It is possible to look at these results from a interesting and different perspective. If we consider that the 2-TLS  are the central system and the photons are the environment then, by solving the exact Hamitonian, we are solving the exact open dynamics of 2-TLS coupled to a bosonic bath. The fact that the correlations are robust with a change in the manifold implies that this family of Bell states has an particular behavior in the presence of dissipation. For the concurrence the robustness can be seen explicitly because the dependence on $n$ is given by:
\begin{equation}
C(t)=\max(0,\Xi(\alpha,t)-f(n)\Omega(\alpha,t)),
\label{concu23+}
\end{equation} where $\Xi(\alpha,t)$ and $\Omega(\alpha,t)$ are oscillating functions of $\alpha$ and $t$ and $f(n)$ is:
\begin{equation}
f(n)=\frac{\sqrt{n(n+1)}}{1+2n},
\label{fn}
\end{equation}
which is a fast growing function that saturates rapidly to $1/2$ ($f'(0)=\infty$ and $f'(\infty)=0$).

%the concurrence is $C(t)=\frac{\sin{2\pi t}^2}{32n^2}+O\left(\frac{1}{n^3}\right)$ **falta dep en $\alpha$***
For $\alpha=0,1$ ($\ket{\psi^+_\alpha}_n=\ket{+-}_n$ or $\ket{-+}_n$), $\Xi(\alpha,t)=\sin^2{2\pi t}=\Omega(\alpha,t)/2$. Therefore the concurrence (Eq. (\ref{concu23+})) is a simple periodic function with oscillations of order $\mathcal{O}(1/n^2)$. It is important to note that the quantum discord for the states  $\alpha=0,1$ is different. The difference stems from the classical correlations (cf Eq.  (\ref{clasiquito})), specifically from the entropies. One can already see it in the elements of the reduced density matrix because $\rho_{22}(t)$ and $\rho_{33}(t)$ are dephased (cf right panel in Fig \ref{0415disc}). Evidently, from the quantum point of view the entanglement is the same because the Hamitonian and the corresponding evolution operator are left unchanged when one permutes the state $\ket{+-}_n$ with $\ket{-+}_n$. On the other side, classical correlations suppose a measurement basis and this election and subsequent minimization induces a difference between the evolution of the two states. The effect is small but important because it is showing that the correlations are not entirely measured by concurrence. As n increases this effect becomes smaller. Please note that here the expression for discord (\ref{discord}) is exact because inequality (\ref{dos}) is satisfied at all times.

For the Bell State of this family $\ket{\psi^+_{1/\sqrt{2}}}_n=\frac{1}{\sqrt{2}}(\ket{+-}_n+\ket{-+}_n)$ $\Xi(\alpha,t)=\cos^2{2\pi t}/4$ and $\Omega(\alpha,t)=\sin^2{2\pi t}/2$ so Eq. (\ref{concu23+}) becomes 
\begin{equation}
C(t)=\max(0,\cos 4 \pi t) + \mathcal{O}\left({1}/{n^2}\right).
\end{equation}
The collapse and revival times correspond to the solutions of equation $\tan{2\pi t}=\pm 1/\sqrt{2f(n)}$ with $f(n)$ given in Eq. (\ref{fn}). As $n\to \infty$, $f(n)\to 1/2$ so the collapse and revival times are $t_i/t_R=(2i-1)/8$, $i=1,2,3,4$  and the ESD lasts $\Delta t=t_R/4$ which is the largest ESD for this initial condition. The quantum discord for the Bell state starts at a maximum $D(0)=1$, vanishes at $t_R/4$ and returns to a maximum at $t_R/2$ presenting two slope discontinuities (cf. Fig. \ref{0313fig23+}).
%*****

We can make some general remarks on the correlations measured by both discord and entanglement in Fig.~\ref{0313fig23+}. Note that the entanglement sudden death (EDS) is accompanied by a discrete vanishing of the quantum discord
%(which is expected since the set of states with zero discord has measure zero and a small perturbation will take a state with zero discord to a state with finite discord CITE) 
but every time the quantum discord vanishes the concurrence vanishes. 
%This is expected since the set of states with discord zero has measure zero.
Furthermore, the discord has oscillations with harmonics of the fundamental Rabi frequency. For the Bell state (see red curve in Fig. \ref{0313fig23+}) the entanglement and discord are in phase and give roughly the same information at least for where the maximum of quantum correlations occurs but for the other initial states they give information about correlations that is completely opposite at a given time (see blue curve in Fig.~\ref{0313fig23+}). The fact that they give opposite information will be seen for other initial conditions.  Consequently, if one asks if the quantum state is correlated or not at $t_R/4$ there is no absolute answer because quantum discord tells us that this is the maximally correlated state while concurrence says that this is a state with no correlations. K. Modi et al., \cite{Modi} proposed a geometrical interpretation of correlations that considers quantum discord as the distance in Hilbert space between the quantum state of the system and the nearest classical pure state and concurrence as the distance to the nearest separable state. From that perspective one can say that in most of the states in the family the separable states are close, in fact, the state is a separable state and at the same time the classical pure states are far in Hilbert space.  Clearly, the number of states with zero concurrence will give us an idea of the size of the set of separable states in the Hilbert space.
%%%%%%%%%%%%%%%%%%%%%%%%%%%%%%%%%%%%%%%%%%%%%%%%%%%%
%%%%%%%%%%%%%%%%%%%%%%%%%%%%%%%%%%%%%%%%%%%%%%%%%%%%
%%%%%%%%%%%%%%%%%%%%%%%%%%%%%%%%%%%%%%%%%%%%%%%%%%%%
%%%%%%%%%%%%%%%%%%%%%%%%%%%%%%%%%%%%%%%%%%%%%%%%%%%%
%%%%%%%%%%%%%%%%%%%%%%%%%%%%%%%%%%%%%%%%%%%%%%%%%%%%
%%%%%%%%%%%%%%%%%%%%%%%%%%%%%%%%%%%%%%%%%%%%%%%%%%%%

\begin{figure}
\includegraphics[width=0.45\textwidth]{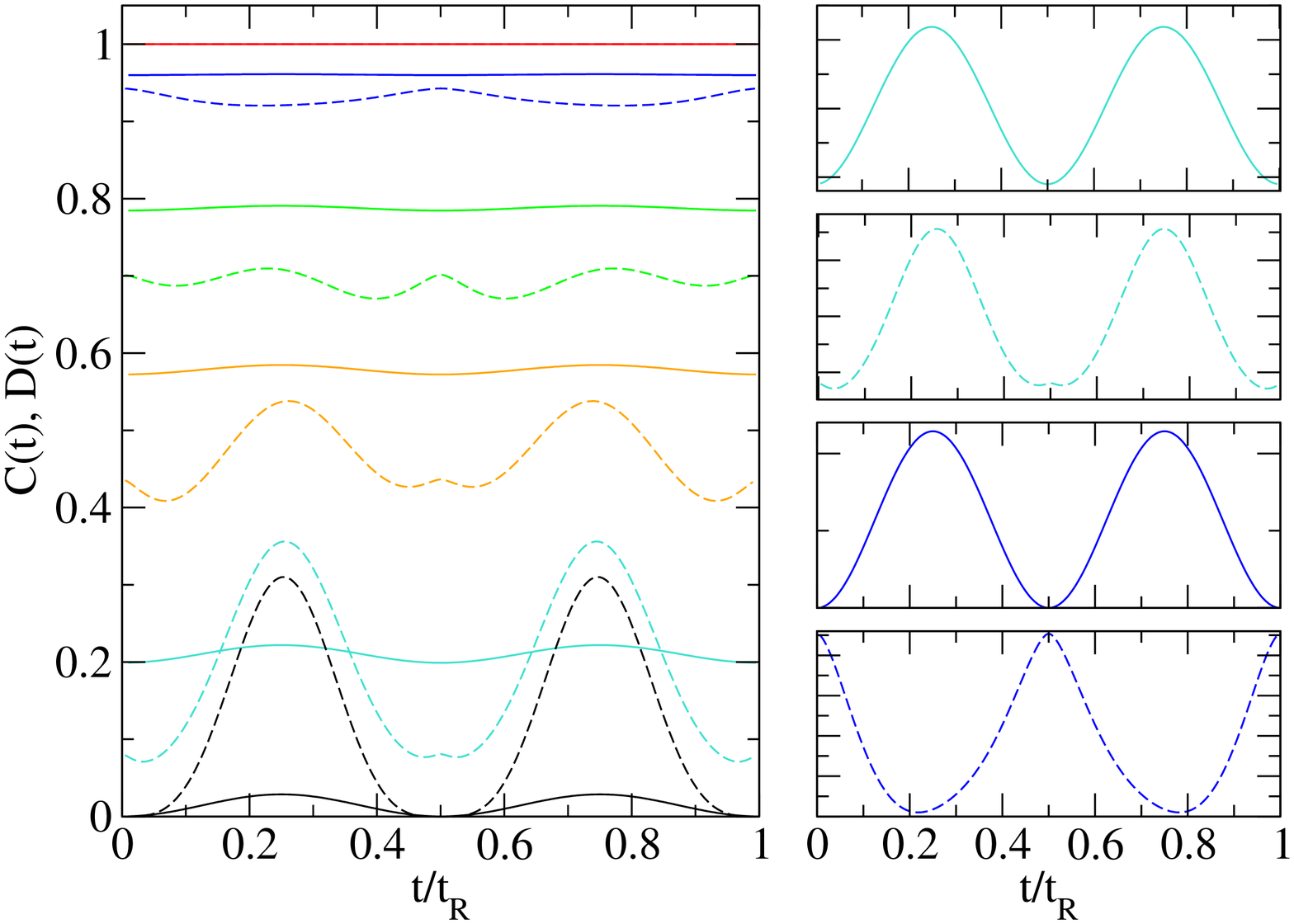}
\caption{\label{0417fig23-_v3}(color online) Left panel: Concurrence (bold lines) and Discord (dashed lines) as a function of time for initial condition $\ket{\psi^-_\alpha}_1$ with $\alpha=0$ (black), $\alpha=0.1$ (turquoise), $\alpha=0.3$ (blue), $\alpha=0.6$ (orange), $\alpha=1/\sqrt{2}$ (red) and $\alpha=0.9$ (green). Right panel: Zoom of oscillations of concurrence and quantum discord for $\alpha=0.1$ (top panels) and $\alpha=0.6$ (bottom panels).}
\end{figure}

The family $\ket{\psi_\alpha^-}_n$ has very different dynamics because in this case the related Bell State $\ket{\psi_{{1}/{\sqrt{2}}}^-}_n=\frac{1}{\sqrt{2}}(\ket{+-}_n-\ket{-+}_n)$ is an exact eigenstate of the Tavis--Cummings Hamiltonian. The concurrence is: 
\begin{equation}
C(t)=g(\alpha)+(\tfrac{1}{2}-f(n))(1-g(\alpha)){\sin^2{2\pi t }},
\end{equation}
with $g(\alpha)$ and $f(n)$ defined in Eqs. (\ref{galpha}) and (\ref{fn}). Therefore, the correlations measured by concurrence present oscillations with amplitude $\tfrac{(1-g(\alpha))}{16n^2}+\mathcal{O}(1/n^3)$ around $g(\alpha)$. The prefactor of the oscillating term, of order $\mathcal{O}(1/n^2)$, is smaller than the constant term $g(\alpha)$, so the entanglement dynamics will be quasi-stationary with a time averaged concurrence that can be tuned by the initial condition (cf. Fig. \ref{0417fig23-_v3}). As $n\to\infty$ the amplitude of oscillations goes to zero. From the point of view of considering the photons as a bath for the 2-TLS, the family $\ket{\psi_\alpha^-}_n$ is also robust against dissipation in the limit $n\to\infty $. For a given manifold $\Lambda_n$ as $\alpha$ goes from $\alpha=0$ to $\alpha=1/\sqrt{2}$ the amplitude of oscillations decreases and the entanglement increases reaching a maximum of $1$. For parameters between $\alpha=1/\sqrt{2}$ and $\alpha=1$ the amplitude of oscillations increases until the entanglement is zero and the oscillations have the largest amplitude.

The dynamics of discord are always given by expression (\ref{discord}) and are also quasi-stationary oscillations (cf Fig. \ref{0417fig23-_v3}). The time averaged discord for a given value of $\alpha$ has the same overall behavior of the quantum correlations measured by concurrence and, as expected for Bell States, both discord and concurrence are equal to one for $\alpha=1/\sqrt{2}$. 
%The dependance of quantum discord is also of order $\mathcal{O}(1/n^2)$
However, the underlying oscillations of both measures have different behaviors. In general, the amplitude of oscillations in discord is larger than the amplitude of oscillations in concurrence and, as for the previous family, the discord often has oscillations with harmonics of the fundamental Rabi frequency.  Furthermore, for some initial conditions  ($\alpha=0.1$, top insets in Fig. \ref{0417fig23-_v3}) both oscillations are in phase but for other ($\alpha=0.6$, bottom insets in Fig. \ref{0417fig23-_v3}) they are in counter phase giving opposite information about the quantum correlations of the state at a given time. We also observe that for some times as concurrence augments discord decreases.  The geometric interpretation of this behavior is that for those states $\rho(t)$ is no longer a separable state and the distance to the closest classical state decreases implying that the classical states are close to the frontier of separable states in the Hilbert space.

The Bell state associated with this initial condition  $\ket{\psi_{1/\sqrt{2}}^-}_n$ belongs to the decoherence free subspace \cite{Eberly2007}. This means that even when the system is open and dissipation is present the entanglement of this state will be one.  Qi-Liang He et al., \cite{He2011} have observed that the discord and entanglement in a similar arrangement but under dissipation have a quasi-stationary oscillatory behavior for some initial conditions related to the family $\ket{\psi_\alpha^-}_n$, we suspect this is due to the quasi-stationary oscillatory exact dynamics reported here. The fact that the concurrence can be tuned by changing the initial condition combined with the possible robustness against dissipation makes these states good candidates for quantum computation purposes.

An initial condition related to the Bell eigenstate  $\ket{\psi_{1/\sqrt{2}}^-}_n$ that is often studied is the Werner state 
\begin{equation}
\rho_W=\alpha\ket{\psi_{1/\sqrt{2}}^-}_n\prescript{}{n}{\bra{\psi_{1/\sqrt{2}}^-}}+\frac{1-\alpha}{4}\mathcal{I}_n,
\end{equation}
with $\mathcal{I}_n$ the identity in the manifold $\Lambda_n$. 
%=\ket{1}_{n-1}\bra{1}_{n-1}+\ket{2}_n\bra{2}_n+\ket{3}_n\bra{3}_n+\ket{4}_{n+1}\bra{4}_{n+1}$. 
This state, proposed by Werner in a historically important article \cite{Werner}, is interesting because for $\alpha=1$ it corresponds to the maximally entangled state, for $\alpha=0$ to the maximally mixed state and for intermediate $\alpha$ to a state that is both entangled and mixed to some degree. The entanglement and quantum discord for this state are initially $C(0)=\max(0,\frac{3\alpha-1}{2})$ and $D(0)=\tfrac{1}{4}((1-\alpha)\log_2(1-\alpha)+(1+3\alpha)\log_2(1+3\alpha)-2(1+\alpha)\log_2(1+\alpha))$. Interestingly, the Werner state does not evolve under the Tavis--Cummings Hamiltonian so we have an example of initial state with exact stationary dynamics were the value of quantum correlations can be tuned monotonically from $0$ to $1$ by changing $\alpha$ from $0$ to $1$ in the initial condition.

%%%%%%%%%%%%%%%%%%%%%%%%%%%%%%%%%%%%%%%%%%%%%%%%%%%%
%%%%%%%%%%%%%%%%%%%%%%%%%%%%%%%%%%%%%%%%%%%%%%%%%%%%
%%%%%%%%%%%%%%%%%%%%%%%%%%%%%%%%%%%%%%%%%%%%%%%%%%%%
%%%%%%%%%%%%%%%%%%%%%%%%%%%%%%%%%%%%%%%%%%%%%%%%%%%%
%%%%%%%%%%%%%%%%%%%%%%%%%%%%%%%%%%%%%%%%%%%%%%%%%%%%
%%%%%%%%%%%%%%%%%%%%%%%%%%%%%%%%%%%%%%%%%%%%%%%%%%%%

\subsection{Family of Bell type States}
\begin{figure}
\includegraphics[width=0.45\textwidth]{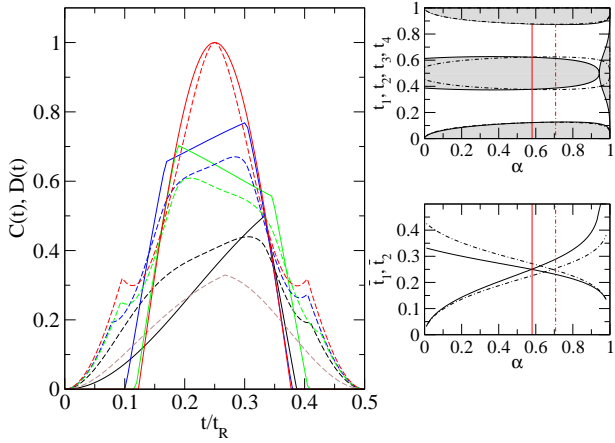}
\caption{\label{0318fig14+_v3}(color online) Left panel: Concurrence (bold lines) and Discord (dashed lines) as a function of time for initial condition $\ket{\phi^+_\alpha}_1$ with $\alpha=0$ (black), $\alpha=0.3$ (blue), $\alpha=\alpha_{B}\simeq0.58$ (red), $\alpha=0.9$ (green) and $\alpha=1$ (brown). Right top panel: Collapse and revival times $t_i/t_R$, $i=1,2,3,4$ as a function of $\alpha$ for $\Lambda_1$ (bold) and $\Lambda_{10}$ (point dashed) Right bottom panel: Discontinuity times $\bar t_i/t_R$, $i=1,2$ as a function of $\alpha$ for $\Lambda_1$ (bold) and $\Lambda_{10}$ (point dashed). The shaded areas in the inset correspond to the ESD and the vertical lines in the right panels correspond to $\alpha_B(1)$ (bold) and $\alpha_B(10)$ (point dashed). }
\end{figure}

The second type of initial conditions we consider are a linear combination of the two other states in the manifold: 
\begin{equation}
%\ket{\phi^\pm_\alpha}=\alpha \ket{1}_{n-1}\pm\sqrt{1-\alpha^2}\ket{4}_{n+1}
\ket{\phi^\pm_\alpha}_n=\alpha \ket{++}_{n-1}\pm\sqrt{1-\alpha^2}\ket{--}_{n+1}.
\label{belltype}
\end{equation}
For $\alpha=0,1$ these correspond respectively to $\ket{--}_{n+1}$ and $\ket{++}_{n-1}$ and  counterintuitively, do not have the same evolution of entanglement. Mathematically this arises from the 
%non-commensurability of the frequencies $\sqrt{n}$ and $\sqrt{n+1}$ in expression \ref{exact} **check ref**. In fact, in a toy model where such dependency on the frequencies is eliminated the concurrence of the two states of the computational basis is the same. 
well known non-linear character of the TC model. On one hand, the evolution of entanglement for the state $\ket{++}_{n-1}$ is always zero. On the other hand, the entanglement of the state $\ket{--}_{n+1}$ starts at a value $C(0)=0$ then augments to a maximum that corresponds to a slope discontinuity at $t_R/4<\bar t_2\leq t_R/2$ and collapses at $t_R/4<t_2<t_R/2$. It then revives at $t_3=t_R-t_2$ to the same value it had at $\bar t_2$ decreasing to $C(t_R)=0$ where it completes a period of evolution. Both the discontinuity times $\bar t_i$, $i=1,2$ ($\bar t_1=0$ for $\ket{--}_{n+1}$) and the collapse and revival times $t_i$, $i=1,2,3,4$ ($t_1=t_4=0$ for $\ket{--}_{n+1}$) are plotted in the right panels of Figure \ref{0318fig14+_v3}. It is straightforward to see that for the initial state $\ket{--}_{n+1}$, $\bar t_2$ is a solution of: 
\begin{equation}
\cos(2\pi t)=-\frac{n}{1+n},
\end{equation}
and $t_2$ is a solution of: 
\begin{equation}
\cos(\pi t)=\sqrt{\Pi(n)},
\end{equation}
with $\Pi(n)=\frac{\sqrt{n (n+1) (2 n+1)^2}} {n+1}-2 n$ . For the manifold $\Lambda_1$ which is the one plotted in Figure  \ref{0318fig14+_v3},  $\bar t_2=1/3$ and $t_2\simeq0.39$.

%Also note that as $n\to \infty$, $\bar t_1=t_2=1/2$.
%\frac{\cos ^{-1}\left(\sqrt{\frac{3}{\sqrt{2}}-2}\right)}{\pi }
The discord for the two basis states $\ket{--}_{n+1}$ and $\ket{++}_{n-1}$ is also different. In contrast to the previous section, the difference does not stem from the classical part but from the classical and quantum contributions taken together in expression (\ref{resta}). As can be seen in Figure \ref{0318fig14+_v3}, the discord for the initial state $\ket{++}_{n-1}$ is non-zero. It oscillates to a maximum near $t_R/4$ and then to zero at $t_R/2$. This is yet another initial condition where there are no correlations measured by concurrence and there are correlations measured by discord. For $\ket{--}_{n+1}$ the discord starts at $D(0)=0$, then augments to a maximum and eventually decreases to zero at $t_R/2$ presenting a slope discontinuity before $t_R/2$. The evolution of quantum discord for $t_R/2<t<t_R$ is symmetric with respect to $t_R/2$. The fact that there is a slope discontinuity in the discord just after the entanglement death and right before the entanglement revival was already encountered for the initial condition $\ket{\psi^+_\alpha}_n$ in the previous section. We conjecture this might be an universal feature.

%(almost in phase, maximum and minimum almost in same place?, when there is a collapse in concurrence there are always to slope discontinuities in the discord?). \\
 %=\cos ^{-1}\left(-1+1/n\right)/2 \pi
 %$\bar t_2=t_R-\bar t_1\geq t_R/2$***
 % collapses \cos^{-1}(\frac{\sqrt{n (n+1) (2 n+1)^2}}{n+1}-2 n)/\pi

We first consider the family $\ket{\phi^+_\alpha}_n$. For  $0<\alpha<1$ there is an initial collapse of the entanglement  $\left[0,t_1\right]$ with $0\leq t_1<t_R/4$, followed by a sudden birth, then followed by an intermediate collapse  $\left[t_2,t_3\right]$ centered at $t_R/2$. The evolution for times $t$ between $t_R/2$ and $t_R$ is symmetric with respect to $t_R/2$. To our understanding this is the first time an initial collapse, with the possibility of returning to a full entangled state, is predicted. While the initial and final collapses, $\left[0,t_1\right]$ and $\left[t_4,t_R\right]$, depend little on the excitation manifold, the intermediate $\left[t_2,t_3\right]$ collapse does depend as can be seen in the top right panel in Figure  \ref{0318fig14+_v3} where we plot the collapse ($t_2$ and $t_4$) and revival ($t_1$ and $t_3$) times for manifolds $\Lambda_1$ and $\Lambda_{10}$. Note that, as $\alpha$ approaches one, the intermediate collapse vanishes, i.e. $t_2=t_3$. This happens at $\alpha_1=2f(n)$ and then, for $\alpha>\alpha_1$, the collapse  time $t_2\to 0$ and the birth time $t_3\to t_R$ until we recover the entanglement dynamics of initial condition$\ket{++}_{n-1}$. Nothing particular happens to discord when $\alpha=\alpha_1$. In general, for the family considered, the discord for $0<\alpha<1$ starts at $D(0)=0$, then increases presenting a slope discontinuity just before the sudden birth of entanglement. In the interval  $\left[t_1,t_2\right]$ it has three behaviours as a function of $\alpha$ that can be seen in Figure \ref{0318fig14+_v3} and then decreases presenting a slope discontinuity right before the concurrence collapses. For times $t_R/2<t<t_R$ the evolution is symmetric with respect to $t_R/2$.

The first interval where the concurrence is different from zero is $\left[t_1,t_2\right]$. For times near $t_R/4$, intermediate $\alpha$ have two slope discontinuities in the concurrence at $\bar t_1$ and $\bar t_2$. The time for which the non analyticities occur for a given $\alpha$ and a given manifold $\Lambda_n$ can be determined analytically by setting $\rho_{11}(t)\rho_{44}(t)=0$ in Eq. (\ref{pmatrixx}) and solving the subsequent  equation. It is straightforward to see that it is the solution of an equation of the form $\cos(2\pi \bar t_i)=P_i(n,\alpha)$ with $P_i(n,\alpha)$ a polynomial function and $i=1,2$. These times are plotted in the right bottom panel in Figure \ref{0318fig14+_v3}. If $P_1(n,\alpha)=P_2(n,\alpha)$ the two slope discontinuities coincide and the entanglement is a smooth function of time, we note $\alpha_B$ the value of $\alpha$ when this happens:
\begin{equation}	
\alpha_B(n)=\sqrt{\frac{n}{1+2n}}.
\end{equation}
The corresponding initial state, the Bell state $\ket{\phi^+_{\alpha_B}}_n$, has maximum quantum discord and maximum entanglement at $t_R/4$. This is an example of a maximally correlated state that depends on the number of initial photons. The result is interesting because starting with an initial uncorrelated state one is able to reach a maximally correlated state.

For $\bar t_1 < t < \bar t_2$, there is a positive slope of concurrence and discord if $\alpha<\alpha_B$ and a negative slope if $\alpha>\alpha_B$ (cf Fig. \ref{0318fig14+_v3}).  The similarity between the evolution of concurrence and discord comes from the whole expression (\ref{resta}) rather than from the classical or quantum parts. The height $C(t_R/4)$ and the slope $C'(t_R/4)$ of the concurrence can be obtained analytically as a function of $n$ and $\alpha$. As $n\to\infty$, $C(t_R/4)\to g(\alpha)$ and $C'(t_R/4)\to0$. This result, combined with the control of the initial ESD by selecting the initial condition $\alpha$ and the control of the intermediate ESD by adjusting $\alpha$ and the number of initial photons in the cavity can be used to control quantum gates where the concurrence is zero or non-zero in certain intervals of time that can controlled and the value of concurrence $C(t_R/4+j t_R/2)$, $j$ an integer, depends on the initial condition.

%From the point of view? of considering the cavity photons? as the environment this means that even in the limit where we have infinite photons the concurrence can still be maximal because 
%Falto: El discord es mas suave a mayor n, en alfa beta tanto discord como la concurrencia son bastante robustos al cambio de variedad (pues, mas o menos porque el alfa beta si depende del acambio de variedad). Para alfa<alfa beta tanto discord como concurrencia la pendiente s baja y se vuelve flat, para alfa > alfa beta tanto discord como concurrencia la pendiente se vuelve flat pero sube. The dependance of concurrence and discord on $n$ is more pronounced close to the maximum and to the discontinuities.\\

%%%%%%%%%%%%%%%%%%%%%%%%%%%%%%%%%%%%%%%%%%%%%%%%%%%%
%%%%%%%%%%%%%%%%%%%%%%%%%%%%%%%%%%%%%%%%%%%%%%%%%%%%
%%%%%%%%%%%%%%%%%%%%%%%%%%%%%%%%%%%%%%%%%%%%%%%%%%%%
%%%%%%%%%%%%%%%%%%%%%%%%%%%%%%%%%%%%%%%%%%%%%%%%%%%%
%%%%%%%%%%%%%%%%%%%%%%%%%%%%%%%%%%%%%%%%%%%%%%%%%%%%
%%%%%%%%%%%%%%%%%%%%%%%%%%%%%%%%%%%%%%%%%%%%%%%%%%%%
\begin{figure}
\includegraphics[width=0.45\textwidth]{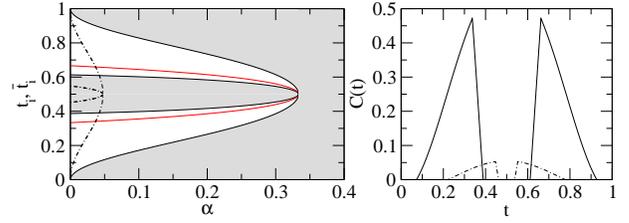}
\caption{\label{0319fig14-andWerner}(color online) Left panel: Collapse and revival times $t_i/t_R$, $i=1,2,3,4$ (black) and discontinuity times $\bar t_i/t_R$, $i=1,2$ (red) a function of $\alpha$ for $\Lambda_1$ (bold) and $\Lambda_{10}$ (point dashed) for initial condition  $\ket{\phi^-_\alpha}_n$. The shaded areas correspond to ESD. Right panel: Concurrence as a function of time for $\alpha=0.02$  for $\Lambda_1$ (bold) and $\Lambda_{10}$ (point dashed)}.
\end{figure}
We now discuss the dynamics of entanglement and quantum discord for initial condition $\ket{\phi^-_\alpha}_n$. There are two distinct evolutions for this initial condition depending on the value of $\alpha$.  For $0<\alpha<\alpha_c(n)$, with\begin{equation}
\alpha_c(n)=\frac{1}{1+2n},
\end{equation}
the concurrence has an initial collapse $\left[0,t_1\right]$ followed by a sudden birth at $t_1$. In the interval $\left[t_1,t_2\right]$ it reaches a maximum that corresponds to a slope discontinuity similar to the situation encountered previously for initial condition $\ket{--}_{n+1}$, collapsing again at $t_2$. Also, like in the previous cases, the evolution is symmetrical with respect to $t_R/2$ for times $t_R/2<t<t_R$.
\begin{figure}
\includegraphics[width=0.45\textwidth]{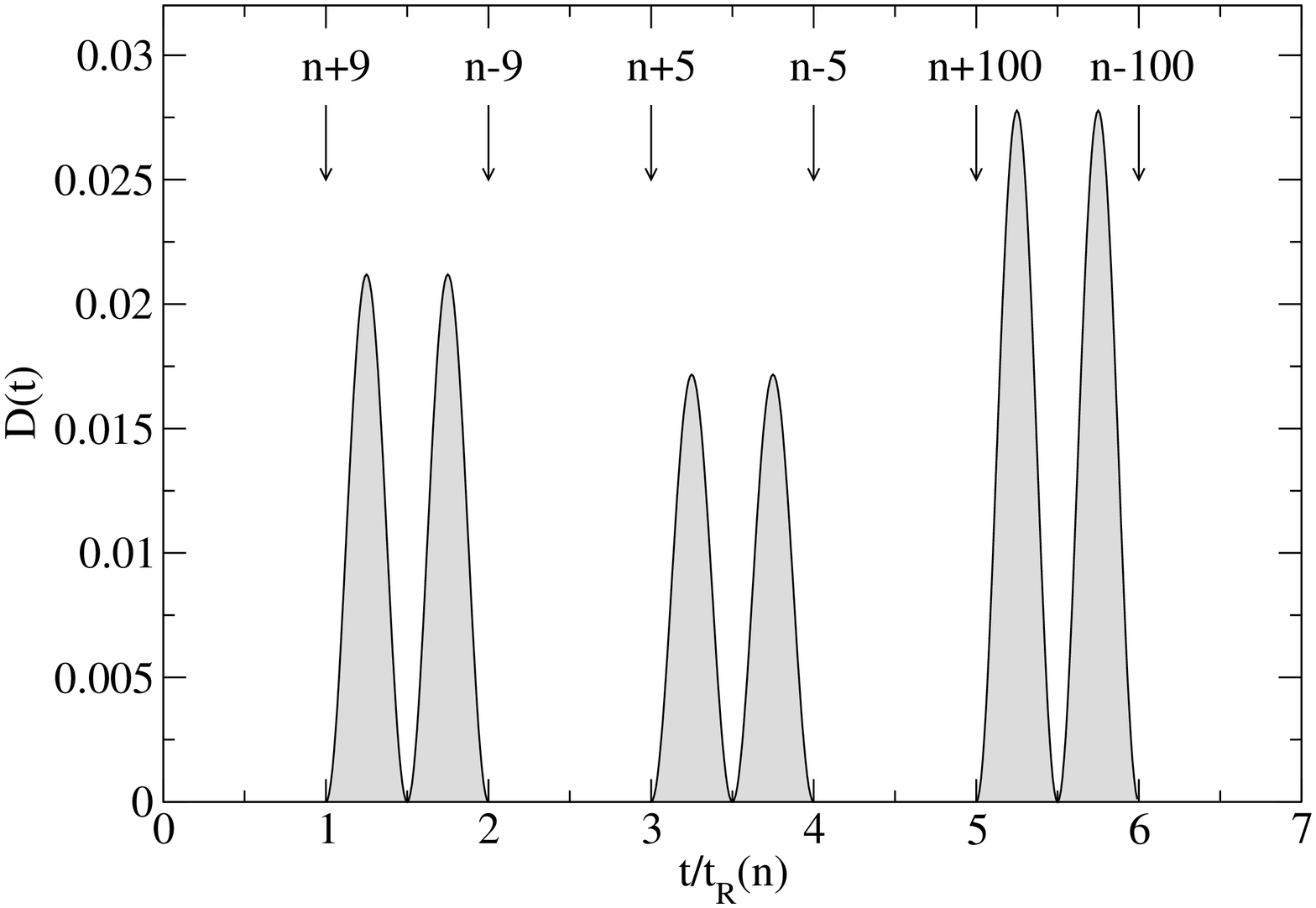}
\caption{\label{gates}Example of a quantum discord gate. Quantum discord as a function of time in units $t_R(n)$ for initial condition $\ket{\phi^-_{\alpha_0}}_1$ with $\alpha_0=\alpha_0(1)\simeq0.82$. The arrows indicate the times at which photons are injected or removed from the cavity.}
\end{figure}
%then by an intermediate collapse $\left[t_2,t_3\right]$  then another birth $\left[t_3,t_4\right]$ and then a final collapse $\left[t_4,t_R\right]$  with a concurrence at $t_R/4$ and $3t_R/4$ that is $g(\alpha)$ for infinite photons in the cavity. 
As $\alpha$ increases the birth intervals $\left[t_1,t_2\right]$ and $\left[t_3,t_4\right]$ become smaller until reaching $\alpha_c$ where the concurrence is zero at all times as for the initial state $\ket{++}_{n-1}$. See left panel in Figure \ref{0319fig14-andWerner} for the times at which the revival/collapse and discontinuities occur in the interval $\left[0,t_R\right]$ and right panel in Figure \ref{0319fig14-andWerner} for a plot of the concurrence as a function of time in manifolds $\Lambda_1$ and $\Lambda_{10}$. From the above description it is straightforward to see that, from the point of view of concurrence, for $\ket{\phi^-_\alpha}_n$ there is no related Bell State because there is no state that is completely entangled. Interestingly, from the point of view of quantum discord there is no related Bell State either.

As in previous initial conditions both quantifiers seem to be uncorrelated because the quantum discord does not change dramatically at $\alpha_c$ and concurrence does not change at all at $\alpha_0$ (see next paragraph). For small values of $\alpha$ in the initial condition the quantum discord starts at $D(0)$ then augments to a maximum decreasing to zero at $t_R/2$ presenting a slope discontinuity just before the ESD similar to initial condition $\ket{\phi^+_\alpha}_n$ and for larger values of $\alpha$ it starts at zero, augments to a maximum near $t_R/4$ and decreases to zero without the slope discontinuity. As $\alpha$ increases the quantum discord is smoother function in time and the maximums start occurring at exactly $t_R/4$.
% NO SIEMPRE QUE HAY ESD HAY DISCONTINUITY EN EL DISCORD
% EL DISCORD NO ES SENSIBLE A LO QUE PASA EN LA CONCURRENCIA
% CUALITATIVO: AL PRINCIPIO (DE ALFA) SE VE DISCONTINUIDAD SIEMPRE QUE HAY SUDDEN DEATH, TAMBIEN SE VEN OTRAS DISCONTINUIDADES UNAS EN EL INITIAL COLLAPSE Y OTRAS EN PARTES RARAS COMO EN EL MAXIMO. A MEDIDA QUE AUMENTA SE PONE SIMETRICO Y LOS MORRITOS SE VAN BAJANDO.

A very surprising feature in this family of initial conditions is that, for each manifold $\Lambda_n$, there is a value of $\alpha$, $\alpha_0(n)$, for which the quantum correlations measured by quantum discord are zero for all times. The value of $\alpha$ for which this happens can be determined by setting the classical correlations equal to the quantum correlations in expression (\ref{resta}) and numerically solving the subsequent transcendental equation. For $\Lambda_1$, $\alpha_0\sim 0.82$ and for $n=\infty$, $\alpha_0=1/\sqrt{2}$ . We propose to use this result, combined with the fact that the state after one complete period of evolution is the same, for quantum computation purposes in an experiment that we call ``Quantum discord gates'', schematized in Figure \ref{gates}. Note that, for this experiment, of the same order of magnitude of the bounded error for discord\cite{Huang}, expression (\ref{discord}) is exact because inequality (\ref{uno}) holds at all times for $1/\sqrt{2}<\alpha<0.82$.  Imagine one starts an experiment with $n$ photons in the cavity and at exactly $\alpha_0(n)$. Then, for the first complete period of evolution $\left[0,t_R\right]$ one has zero quantum discord. Eventually, at $t=t_R$, one injects $m$ photons. Clearly, since $\alpha_0(n+m)\neq\alpha_0(n)$ the discord will be non-zero in the interval $\left[t_R,2t_R\right]$. Evidently, this process can be repeated at any multiple of $t_R$ and, this way, by injecting or extracting photons at certain times one has zero or non-zero discord. From what was just explained, this experiment uses discord to count the photons in the cavity and 
% because if the number of photons is greater than one then one has a non-zero measurement of quantum discord so the quantum discord is actually counting the photons in the cavity.
%***CAMBIAR*** Please note that the quantum discord is actually counting the photons in the cavity. If the number of photons is greater than one, then one has/ THERE IS *** a non-zero measurement of quantum discord.
this is an important result of our work. Returning to the interpretations of the photons as the environment for the 2-TLS and comparing initial conditions $\ket{\phi^-_{\alpha_0}}_n$ and $\ket{\phi^-_{\alpha_0}}_{m}$,
then another conclusion from our work is that initial correlations with the environment have great impact on dynamics of quantum discord. In fact, the reduced 2-TLS density matrix of such initial conditions are indistinguisable
% and equal to: 
%\[ \left( \begin{array}{cc}
%\alpha^2 & \alpha\sqrt{1-\alpha^2}  \\
%\alpha\sqrt{1-\alpha^2}  & 1-\alpha^2  \\
 %\end{array} \right)\] 
but they evolve to completely different discord because one evolves to finite discord while the other initially equivalent condition evolves to an uncorrelated state. 

\subsection{Ali states}
\begin{figure}
\includegraphics[width=0.45\textwidth]{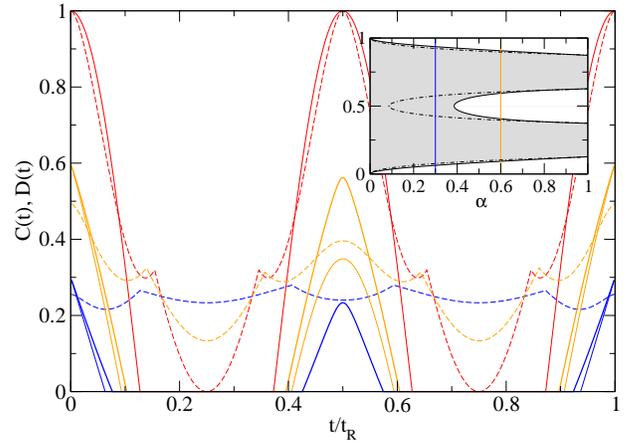}
\caption{\label{0318figali2a}(color online) Concurrence (bold lines) and Discord (dashed lines) as a function of time for initial condition $\rho_A$ with $\alpha=0.3$ (blue), $\alpha=0.6$ (orange) and $\alpha=1$ (red) for manifolds $\Lambda_1$ and $\Lambda_{10}$. The thicker lines for correspond to $\Lambda_{10}$. The inset shows the dependence of collapse and revival times $t_i/t_R$ $i=1,2,3,4$ as a function of $\alpha$ for $\Lambda_1$ (bold) and $\Lambda_{10}$ (point dashed). Here $\alpha_A(1)\simeq0.38$ and $\alpha_A(10)\simeq0.09$. The shaded area corresponds to the ESD and the vertical lines correspond to the values of $\alpha$ considered in the big figure.}
\end{figure}
%%%%%%%%%%%%%%%%%%%%%%%%%%%%%%%%%%%%%%%%%%%%%%%%%%%%
%%%%%%%%%%%%%%%%%%%%%%%%%%%%%%%%%%%%%%%%%%%%%%%%%%%%
%%%%%%%%%%%%%%%%%%%%%%%%%%%%%%%%%%%%%%%%%%%%%%%%%%%%
%%%%%%%%%%%%%%%%%%%%%%%%%%%%%%%%%%%%%%%%%%%%%%%%%%%%
%%%%%%%%%%%%%%%%%%%%%%%%%%%%%%%%%%%%%%%%%%%%%%%%%%%%
%%%%%%%%%%%%%%%%%%%%%%%%%%%%%%%%%%%%%%%%%%%%%%%%%%%%
A mixed initial condition related to $\ket{\psi^\pm_\alpha}_n$  and $\ket{\phi^\pm_\alpha}_n$  that is similar to one studied by Ali et al., \cite{Ali} is:
%$\rho_{A}^{\ket{1}}=\alpha\ket{\psi^{+}_{1/\sqrt{2}}}\bra{\psi^{+}_{1/\sqrt{2}}}+(1-\alpha)\ket{1}_{n-1}\bra{1}_{n-1} $. Since the entanglement and discord dynamics of the states $\ket{1}_{n-1}$ and $\ket{4}_{n+1}$ are different, we consider two families of initial conditions:
\begin{equation}
\rho_{A}=\alpha\ket{\psi^+_{1/\sqrt{2}}}_n\prescript{}{n}{\bra{\psi^+_{1/\sqrt{2}}}}+(1-\alpha)\ket{++}_{ n-1}\prescript{}{n-1}{\bra{++}}. \\
\end{equation}
%\begin{align}
%\rho_{A}=&\alpha\ket{\psi^+_{1/\sqrt{2}}}\bra{\psi^+_{1/\sqrt{2}}}+(1-\alpha)\ket{1}_{ n-1}\bra{1}_{n-1} \\
%\rho_{A}^{\ket{4}}= &\alpha\ket{\psi^+_{1/\sqrt{2}}}\bra{\psi^+_{1/\sqrt{2}}}+(1-\alpha)\ket{4}_{ n+1}\bra{4}_{n+1} 
%\end{align}
%Results for these two initial conditions ***repeticion initial conditions*** can be seen in Figs. \ref{0318figali2a} and \ref{0318figali2b}. 
For this family of initial conditions  there are two distinct behaviors of entanglement determined by:
\begin{equation}
\alpha_A(n)=\left[1+\frac{1+2n}{4f(n)}\right]^{-1}.
\end{equation}  
%$ \prescript{14}{2}{\mathbf{C}} $
%$ \prescript{}{2}{\bra{C}} $
If $\alpha_A<\alpha\leq 1$ the entanglement dynamics resemble the evolution of the maximally entangled state $\ket{\psi^+_{1/\sqrt{2}}}_n$ where concurrence starts at a maximum $g(\alpha)$ (cf Eq. (\ref{galpha})), then an ESD occurs at $0<t_1<t_R/4$ and a sudden revival at $t_R/4<t_2<t_R/2$ augmenting to the same value $g(\alpha)$ at $t_R/2$. In general, the collapse and revival times satisfy $t_1(n)\lesssim t_1(n+1)$, $t_2(n)>t_2(n+1)$, $t_3(n)<t_3(n+1)$ and $t_4(n)\gtrsim t_4(n+1)$ (cf inset of Fig. \ref{0318figali2a}), and as $\alpha\to1$ they are not sensible to the change of manifold. On the other hand, if $0\leq\alpha<\alpha_A$ the evolution is different from all the previous evolutions because for a given $\alpha$, the dependence of entanglement on the number of photons in the cavity is counterintuitive. To illustrate this point, lets examine the concurrence at $t_R/2$. It is straightforward to see that:
 \be
 C(t_R/2)=\max\left[0,\alpha-4(\alpha-1)\frac{f(n)}{1+2n}\right].
 \label{ctr2}
 \ee 
The dependence on $n$ in the above expression implies that as one augments the number of photons in the cavity there is a sudden birth of concurrence at $t_R/2$.
% and the concurrence grows as $n$ grows. 
%and as we keep augmenting the entanglement is larger and larger. 
From the point of view of considering the photons as the environment for the 2-TLS this result  is interesting because increasing the number of photons, i.e. attaining the photon bath limit, results in a more entangled state at certain times.  In Figure \ref{0318figali2a} we plot the concurrence for $\alpha=0.3$ for $\Lambda_1$ and $\Lambda_{10}$ (blue curves) to observe the sudden birth at $t_R/2$ when there are more photons and the concurrence for $\alpha=0.6$ for $\Lambda_1$ and $\Lambda_{10}$ (orange curves) to observe how the entanglement at $t_R/2$ increases as the excitation manifold increases. The counterintuitive dependence on the number of photons can be used to count photons. To clarify this point lets compare the entanglement evolution for different initial conditions (different $\alpha$) for $n$ and $n+1$ photons in the cavity. If  $0\leq\alpha<\alpha_A(n+1)$ then there is no birth at $t_R/2$ regardless of the number of photons. If $\alpha_A(n+1)<\alpha<\alpha_A(n)$ there is birth if there are $n+1$ photons but there is no birth if there are $n$ photons (this is shown in Figure \ref{0318figali2a} but with a difference of 9 photons). Finally, if $\alpha>\alpha_A(n)$ then the entanglement at $t_R/2$ for $n+1$ photons is larger than for $n$ photons.

\begin{figure}
\includegraphics[width=0.45\textwidth]{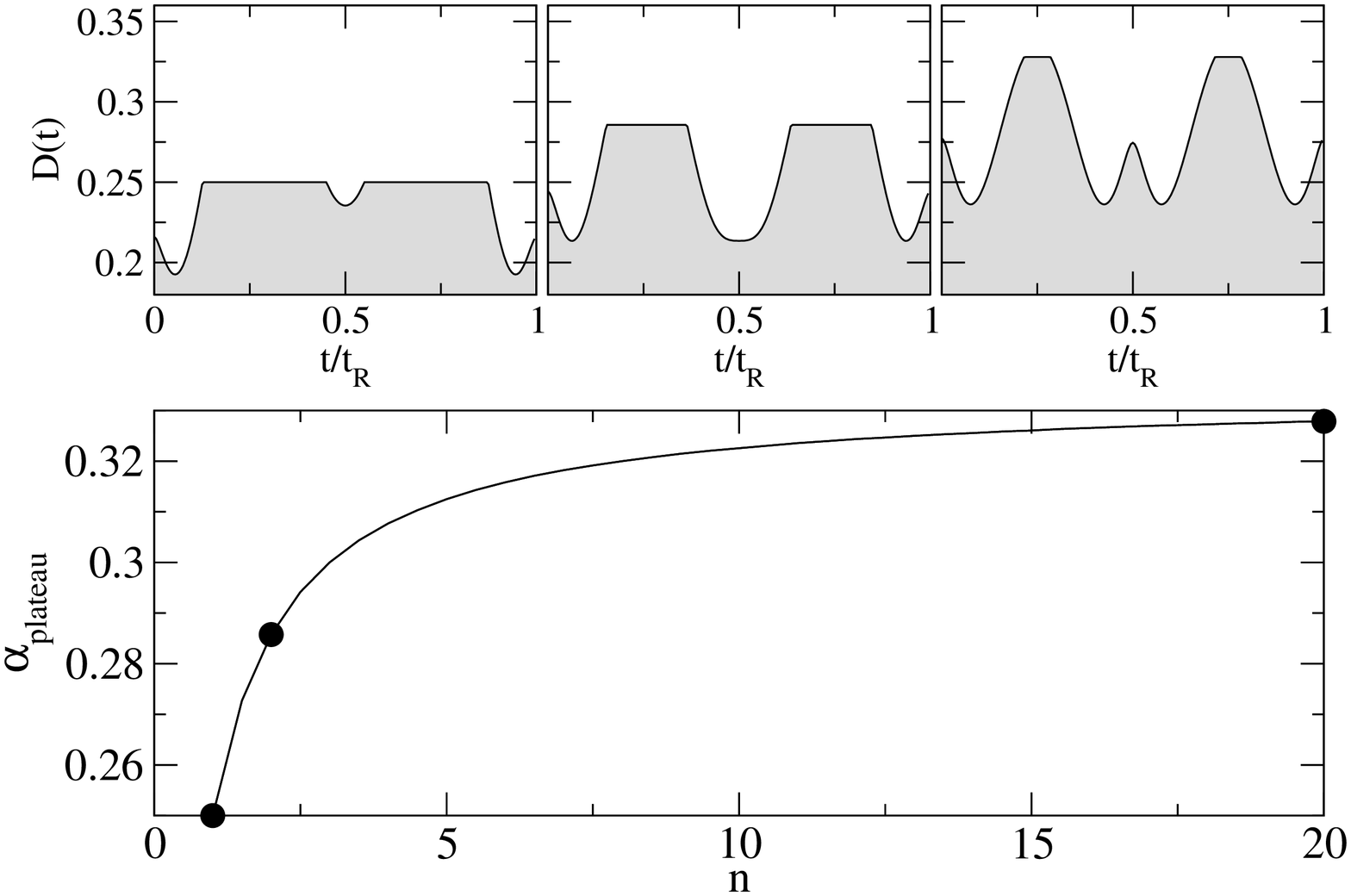}
\caption{\label{0525plateau}Top panels: Discord as a function of time for $\alpha_{plateau}(n)$ for $\Lambda_1$ (left), $\Lambda_2$ (middle) and $\Lambda_{20}$ (right). Bottom panel:  $\alpha_{plateau}(n)$ as a function of $n$.}
\end{figure}

%**in the limit $t\to\infty$ $C(t_R/2)\to\alpha$ **OJO PERO ALFA MENOR QUE ALFA CRITICO Y ALFA CRITICO ES MUY PEQUENO SI HAY MUCHOS FOTONES ENTONCES MEJOR NO DECIR ESTO***. 
% In Section ***REF** we observed that the entanglement of $\ket{1}_{n-1}$ was zero. Here, ***ES FACIL VER ESTO PORQUE*** limit $\alpha\to0$ corresponds to the situation where there is no entanglement at $t_R/2$, i.e. $t_2=t_3$,  $t_1\to0$ and $t_4\to t_R$.\\
For a given number of photons the quantum discord for this initial condition has different evolutions. For $\alpha=0$ it was already described (cf black curve in Fig. \ref{0318fig14+_v3}). For small $\alpha$ it has the shape of a garland with discontinuities at the maximums and smooth minimums as can be seen in the blue curve in Figure \ref{0318figali2a}. As $\alpha$ increases there is a maximum that appears at $t_R/2$ and at the same time the minimums near $t_R/4$ and $3t_R/4$ go to zero (see orange curve in Fig. \ref{0318figali2a}). Finally as $\alpha\to 1$  the maximum at $t_R/2$ is the same as the initial discord and we obtain the known result for Bell where the discontinuities in the discord occur just after the entanglement sudden death and right before the entanglement sudden birth. An new unreported result for the quantum discord is that there is a value of $\alpha$ for each value of $n$ which there is a plateau near $t_R/4$ and near $3t_R/4$ (cf Fig. \ref{0525plateau}).  For one photon in the cavity, $\alpha_{plateau}(1)=\frac{1}{4}$,  and for infinite photons $\alpha_{plateau} (\infty)=\frac{1}{3}$. The discord $D(\alpha_{plateau})$ is $\alpha_{plateau}$ and the width of the plateau decreases as $n$ increases. In Fig. \ref{0525plateau} we plot $\alpha_{plateau}$ as a function of $n$ and the discord for the excitation manifolds $\Lambda_1$,  $\Lambda_2$ and $\Lambda_{20}$.\\
%$1<n<2$, for $n=2$, for $n>2$ 
%***mencionar esta cosa o dejar asi, puede ser contraproducente porque es como mostrar un ncritico para el discord for a given alfa, too much information***. HACER COMENTARIO DE QUE ESTE ES UN RESULTADO INTERESANTISIMO DEL DISCORD Y QUE NO HA SIDO REPORTADO, O SI HA SIDO REPORTADO DECIR QUE ESTO PUEDE SER LA EXPLICACION DE TAL Y TAL COSA.\\
As a final remark for this initial condition, we note that for some initial conditions $\alpha$ and number of photons $n$  the information of the correlations given by concurrence and quantum discord is completely opposite. We have both the situation where entanglement is zero and discord is maximum and also the situation where one augments as the other decreases. 
\section{Discussion}
In this paper we have studied the dynamics of entanglement and quantum discord in the exact Tavis--Cummings Hamitonian. We concluded that some of the dynamical features of discord attributed to the Markovian or non-Markovian nature of the environment in the open system scheme are already contained in the non linear exact dynamics of the TC model and in the choice of initial conditions. In particular, we showed that the discontinuities Maziero et al., \cite{Serra} observed in discord were a direct consequence of the minimization of the entropies. We furthermore demonstrated that the stationary asymptotic dynamics reported by Qi-Liang et al., \cite{He2011} are related to the choice of initial conditions.  
%We also observerd exponential decay and the asymptotic vanishing of*. exponential decay and only-asymptotic vanishing of quantum discord, quasi-stationariedad del discord y entanglemtne para condiciones iniciales relacionadas con eigenstate. 
This simple model reveals that the assumption of an environment is not essential. 
Regarding the entanglement, measured by the concurrence, we exhaustively studied its dynamics, in particular the well known EDS phenomena. In this context, we described the death and revival times as a function of all the the meaningful parameters in the model. These predictions on discord and entanglement will give new insights on the current debate on quantum correlations and the true influence of dissipation and non markovianity on their dynamics. One question still unanswered is what is the real role of dissipation and markovian character of the bath.
%With respect to previous studies, ... this model has several important characteristics in the classical case i) It presents ...ii) It presents a large ...
%We previously reported
%We also stress once again that
%We investigated *** by means of *** 
%We showed that
%In particular, in the
%Additionaly, we showed that
%Moreover
%One would therefore obtain
%This predictions will hopefully give new insights on the current theoretical debate on the nature of 
%Moreover, the present study is quite important, in our opinion, since it opens 
%We expect that this is
%BREUER
%I%n this paper we have studied ***  which was introduced previously
%Our results demonstrate the *** due to 
%For these *** states we have shown 	
%For known *** In this way, the locally non-interacting 
%We have furthermore demosntrated that **+ is not necessary for 
%Finally the phenomena *** has been explained by means of ***
%The general model reveals that the assumption *** are essential in order to obtain this condition
%PRA NUEVA
%Several other ideas will be further investigated in the future
%Given a , it would be interesting to***
%It is also worth further research to characterize
%Last, it is worth studying whether last

The model reveals that the dynamics of correlations depend crucially on the 2-TLS + field initial conditions and are, by no means, trivial. We show that there exist initial conditions where both measures give opposite information: on one hand quantum discord suggests that it is a maximally correlated state and on the other hand concurrence suggest that it is an uncorrelated state. There are also states where both measures are in phase and give roughly the same information for correlations. Interestingly, we find that there are states that are indistinguishable from the point of view of entanglement but distinguishable from the point of view of quantum discord because classical correlations suppose a measurement basis and this election and subsequent minimization induces a difference between the evolution of the two states. Finally, critical behaviours for both quantifiers usually occur in disjoint regions of the parameters space allowing the possibility of building devices where one can control one or the other independently. According to the Modi et al., \cite{Modi} interpretation of quantum correlations, the results just mentioned have implications for the geometry of Hilbert space, for example for the size of the separable states. We leave these as directions for future work.
%ii) where both give the same information iii) where there are two states indistinguishibles by concurrence and with different discord. Finally, Comportamiento critico ocurre en distintas regiones del espacio de parametros lo cual permite hacer control para uno o otro. he study of the geometry of Hilbert space in the TV will be further investigated in the future. 
%It presents ***when they give opposite information for example one is maximum meaning that there and one is zero. i) Sometimes they are in phase or counterphase. iii) 
%ambien es interesante el caso en el que tenemos dos estados indistiguibles por entrelazamiento pero con diferente discordia y que tiene que ver con las correlaciones clasicas. Copy paste: Evidently, from the quantum point of view the entanglement is the same because the Hamitonian and the corresponding evolution operator are left unchanged when one permutes the state $\ket{+-}_n$ with $\ket{-+}_n$. On the other side, . The effect is small but important because it is showing that the correlations are not entirely measured by concurrence. .Eso muestra que la geometria del espacio de estados, aun para este sistema tan sencillo es no trivial y falta mucho por entender. For some initial conditions we found that We shoed that . T\\

The theoretical findings are of direct practical relevance. Our results predict i) Robust maximally entangled states with entanglement that is almost independent on the manifold ii) States whose entanglement and quantum discord can be tuned from zero to one by varying a parameter in the initial condition iii) States that have the possibility of returning to a fully entangled state after an initial ESD where we can control the entanglement by setting the number of initial photons and the parameter in the initial condition iv) States that counterintuitively have an entanglement birth as one augments the excitation number. We have furthermore proposed an experiment called Quantum discord gates where by injecting or extracting photons at certain times the quantum state has zero or non-zero discord. This makes use of a surprising result where we find that for some initial conditions the quantum discord is exactly zero. Given the marginal character of states with zero discord this result is not only completely counterintuitive but is also useful as a way to count photons present in a cavity.

\begin{acknowledgments}
This work was supported by the Vicerrector\'\i a de Investigaci\'on of the Universidad Antonio Nari\~no, Colombia under project number 2011282, Comit\'e para el Desarrollo de la Investigaci\'on(CODI) of the Universidad de Antioquia, Colombia under contract number E01620, Estrategia de Sostenibilidad del Grupo de F\'\i sica At\'omica y Molecular and by the Departamento Administrativo de Ciencia, Tecnolog\'\i a e Innovaci\'on (COLCIENCIAS) of Colombia under grant number 111556934912.
\end{acknowledgments}


\begin{thebibliography}{99}
\bibitem{EPR} A. Einstein, B. Podolsky, and N. Rosen, Phys. Rev. {\bf 47}, 777 (1935).

\bibitem{Bohr} N. Bohr, Phys. Rev. {\bf 48}, 696 (1935).

\bibitem{Olival} O. Freire, {\it The Quantum Dissidents: Rebuilding the Foundations of Quantum Mechanics (1950-1990)} (Springer-Verlag, Berlin, 2015).

\bibitem{Schrodinger} E. Schr\"odinger, Math. Proc. Cambridge {\bf 31}, 555 (1935).

\bibitem{HorodeckiRev} R. Horodecki, P. Horodecki, M. Horodecki, and K. Horodecki, Rev. Mod. Phys. {\bf 81}, 865 (2009).

\bibitem{EntanglementRev}  O. G\"uhne and  G. T\'oth, Phys. Rep. {\bf 474}, 1 (2009).

\bibitem{Werner} R. F. Werner, Phys. Rev. A {\bf 40}, 4277 (1989).

\bibitem{Bell} J.S. Bell, Physics {\bf 1}, 195 (1964). Reprinted in J. S. Bell, {\it Speakable and Unspeakable in Quantum Mechanics} (Cambridge University Press, Cambridge, 2004).

\bibitem{Aspect} A. Aspect, P. Grangier, and G. Roger, Phys. Rev. Lett. {\bf 49}, 91 (1982); A. Aspect, J. Dalibard, and G. Roger, Phys. Rev. Lett. {\bf 49}, 1804 (1982).  

\bibitem{Zeilinger} G. Weihs, T. Jennewein, C. Simon, H. Weinfurter, and A. Zeilinger, Phys. Rev. Lett. {\bf 81}, 5039 (1998); M. Giustina, A. Mech,	S. Ramelow, B. Wittmann, J. Kofler, J. Beyer, A. Lita, B. Calkins, T. Gerrits, S. W. Nam,	R. Ursin and A. Zeilinger, Nature {\bf 497}, 227 (2013).

\bibitem{Gisin} W. Tittel, J. Brendel, H. Zbinden, and N. Gisin, Phys. Rev. Lett. {\bf 81}, 3563 (1998); D. Salart, A. Baas, J. A. W. van Houwelingen, N. Gisin, and H. Zbinden, Phys. Rev. Lett. {\bf 100}, 220404 (2008).

\bibitem{Nielsen} M. A. Nielsen, and I. L. Chuang, {\it Quantum Computation and Quantum Information: 10th Anniversary Edition} (Cambridge University Press, Cambridge, 2010).

\bibitem{Bennet} C. H. Bennett, D. P. DiVincenzo, C. A. Fuchs, T. Mor, E. Rains, P. W. Shor, J. A. Smolin, and W. K. Wootters, Phys. Rev. A {\bf 59}, 1070 (1999).

\bibitem{Bartlett} G. J. Pryde, J. L. O'Brien, A. G. White, and S. D. Bartlett, Phys. Rev. Lett. {\bf 94}, 220406 (2005).

\bibitem{Braunstein} S. L. Braunstein, C. M. Caves, R. Jozsa, N. Linden, S. Popescu, and R. Schack, Phys. Rev. Lett. {\bf 83}, 1054 (1999).

\bibitem{Lanyon} B. P. Lanyon, M. Barbieri, M. P. Almeida, and A. G. White, Phys. Rev. Lett. {\bf 101}, 200501 (2008).
%First paragraph

\bibitem{Zurek} H. Ollivier and W. H. Zurek, Phys. Rev. Lett. {\bf 88}, 017901 (2001).

\bibitem{Vedral} L. Henderson, and V. Vedral, J. Phys. A: Math. Gen. {\bf 34}, 6899 (2001).

\bibitem{Winter} D. Cavalcanti, L. Aolita, S. Boixo, K. Modi, M. Piani, and A. Winter, Phys. Rev. A {\bf 83}, 032324 (2011).

\bibitem{Datta} V. Madhok and A. Datta, Phys. Rev. A {\bf 83}, 032323 (2011).

\bibitem{Dakic} B. Daki\'c, Y. O. Lipp,	X. Ma, M. Ringbauer, S. Kropatschek, S. Barz,	T. Paterek,	V. Vedral, A. Zeilinger,	C. Brukner and P. Walther, Nat. Phys. {\bf 8}, 666 (2012).

\bibitem{Piani} T. K. Chuan, J. Maillard, K. Modi, T. Paterek, M. Paternostro, and M. Piani, Phys. Rev. Lett. {\bf 109}, 070501 (2012).

\bibitem{Pirandola} S. Pirandola, Sci. Rep. {\bf 4}, 6956 (2014).

\bibitem{Adesso2014} D. Girolami, A. M. Souza, V. Giovannetti, T. Tufarelli, J. G. Filgueiras, R. S. Sarthour, D. O. Soares-Pinto, I. S. Oliveira, and Gerardo Adesso, Phys. Rev. Lett. {\bf 112}, 210401 (2014).

\bibitem{Modi} K. Modi, T. Paterek, W. Son, V. Vedral, and M. Williamson, Phys. Rev. Lett. {\bf 104}, 080501 (2010).

\bibitem{Acin} A. Ferraro, L. Aolita, D. Cavalcanti, F. M. Cucchietti, and A. Ac\'in, Phys. Rev. A {\bf 81}, 052318 (2010).

\bibitem{Modi2012} K. Modi, A. Brodutch, H. Cable, T. Paterek, and V. Vedral, Rev. Mod. Phys. {\bf 84}, 1655 (2012).

\bibitem{Streltsov} A. Streltsov, H. Kampermann, and D. Bruß, Phys. Rev. Lett. {\bf 106}, 160401 (2011).

\bibitem{Piani2012} M. Piani, S. Gharibian, G. Adesso, J. Calsamiglia, P. Horodecki, and A. Winter, Phys. Rev. Lett. {\bf 106}, 220403 (2011).
%Second paragraph

\bibitem{Buchleitner} F. Mintert, A. R.R. Carvalho, M. Ku\'{s}, and A. Buchleitner, Phys. Rep. {\bf 415}, 207 (2005).

\bibitem{YuEberly} T. Yu, and J. H. Eberly, Phys. Rev. Lett. {\bf 93}, 140404 (2004); T. Yu, and J. H. Eberly, Science {\bf 323}, 598 (2009).

\bibitem{Davidovich} M. P. Almeida, F. de Melo, M. Hor-Meyll, A. Salles, S. P. Walborn, P. H. Souto Ribeiro, and L. Davidovich, Science {\bf 316}, 579 (2007).

\bibitem{Kimble} J. Laurat, K. S. Choi, H. Deng, C. W. Chou, and H. J. Kimble, Phys. Rev. Lett. {\bf 99}, 180504 (2007).

\bibitem{Fanchini2009} T. Werlang, S. Souza, F. F. Fanchini, and C. J. Villas Boas, Phys. Rev. A {\bf 80}, 024103 (2009).

\bibitem{Celeri} J. Maziero, L. C. C\'eleri, R. M. Serra, and V. Vedral, Phys. Rev. A {\bf 80}, 044102 (2009).

\bibitem{Serra} J. Maziero, T. Werlang, F. F. Fanchini, L. C. C\'eleri, and R. M. Serra, Phys. Rev. A {\bf 81}, 022116 (2010).

\bibitem{Mazzola} L. Mazzola, J. Piilo, and S. Maniscalco, Phys. Rev. Lett. {\bf 104}, 200401 (2010).

\bibitem{Fanchini2010} F. F. Fanchini, T. Werlang, C. A. Brasil, L. G. E. Arruda, and A. O. Caldeira, Phys. Rev. A {\bf 81}, 052107 (2010).
%TC Model

\bibitem{TC} M. Tavis, and F. W. Cummings, Phys. Rev. {\bf 170}, 379 (1968).

\bibitem{JC} E. T. Jaynes, and F. W. Cummings, Proc. IEEE {\bf 51}, 89 (1963).

\bibitem{ReviewJC} B. W. Shore and P. L. Knight, J. Mod. Opt. {\bf 40}, 1195 (1993).

\bibitem{Walls-Milburn} D. F. Walls, and G. J. Milburn, {\it Quantum Optics, Second Ed.} (Springer-Verlag, Berlin, 2010).

\bibitem{Special} See the papers in the special issue J. Phys. B: At. Mol. Opt. Phys. {\bf 46}, 1 (2013), commemorating fifty years in the JC physics.

\bibitem{Haroche} J. M. Raimond, M. Brune, and S. Haroche, Rev. Mod. Phys. {\bf 73}, 565 (2001).

\bibitem{Walther} H. Walther, B. T. H. Varcoe, B.-G. Englert, and T. Becker, Rep. Prog. Phys. {\bf 69}, 1325 (2006).

\bibitem{Wineland1996}  D. M. Meekhof,  C. Monroe, B. E. King, W. M. Itano, and D. J. Wineland, Phys. Rev. Lett. {\bf 76}, 1796 (1996).

\bibitem{WinelandRMP} D. Leibfried, R. Blatt, C. Monroe, and D. Wineland, Rev. Mod. Phys. {\bf 75}, 281 (2003).

\bibitem{Nico} N. Quesada and A. Sanpera, J. Phys. B: At. Mol. Opt. Phys. {\bf 46}, 224002 (2013).

\bibitem{Azuma} H. Azuma, Prog. Theor. Phys. {\bf 126}, 369 (2011).

\bibitem{Molmer} B. Mischuck and K. Molmer, Phys. Rev. A {\bf 87}, 022341 (2013).

\bibitem{Schoelkopf} A. Blais, R.-S. Huang, A. Wallraff, S. M. Girvin, and R. J. Schoelkopf, Phys. Rev. A {\bf 69}, 062320 (2004).

\bibitem{Schoelkopf2004} A. Wallraff, D. I. Schuster, A. Blais, L. Frunzio, R.-S. Huang, J. Majer, S. Kumar, S. M. Girvin, and R. J. Schoelkopf, Nature {\bf 431}, 162 (2004).

\bibitem{Mooij} I. Chiorescu, P. Bertet, K. Semba, Y. Nakamura, C. J. P. M. Harmans, and J. E. Mooij, Nature {\bf 431}, 159 (2004).

\bibitem{Takayanagi} J. Johansson, S. Saito, T. Meno, H. Nakano, M. Ueda, K. Semba, and H. Takayanagi, Phys. Rev. Lett. {\bf 96}, 127006 (2006).

\bibitem{KiraKoch} M. Kira, and S. W. Koch, {\it Semiconductor Quantum Optics} (Cambridge University Press, Cambridge 2011).

\bibitem{Forchel} J. P. Reithmaier, G. Sek, A. L\"offler, C. Hofmann, S. Kuhn, S. Reitzenstein, L. V. Keldysh, V. D. Kulakovskii, T. L. Reinecke, and A. Forchel, Nature {\bf 432}, 197 (2004).

\bibitem{Deppe} T. Yoshie, A. Scherer, J. Hendrickson, G. Khitrova, H. M. Gibbs, G. Rupper, C. Ell, O. B. Shchekin, and D. G. Deppe, Nature {\bf 432}, 200 (2004).

\bibitem{Elena} E. del Valle, {\it Microcavity Quantum Electrodynamics} (VDM Verlag, 2010).

\bibitem{Elena2012} F. P. Laussy, E. del Valle, M. Schrapp, A. Laucht, and J. J. Finley, J. Nanophoton. {\bf 6}, 061803 (2012).

\bibitem{us} C. A. Vera, N. Quesada M., H. Vinck-Posada, B. A. Rodriguez, J. Phys.: Condens. Matt. {\bf 21}, 395603 (2009).

\bibitem{Marquardt} M. Aspelmeyer, T. J. Kippenberg, and F. Marquardt, Rev. Mod. Phys. {\bf 86}, 1391 (2014).

\bibitem{Milburn-Woolley} G. J. Milburn and M. J. Woolley, Acta Phys. Slovaca {\bf 61}, 483 (2011).

\bibitem{Kim2002} M. S. Kim, Jinhyoung Lee, D. Ahn, and P. L. Knight, Phys. Rev. A {\bf 65}, 040101 (2002).

\bibitem{Puri} R. R. Puri, {\it Mathematical Methods of Quantum Optics} (Springer-Verlag, Berlin, 2001).

%Ultimas referencias (X-states, ESD, etc)

\bibitem{Eberly2007} T. Yu and J. H. Eberly, Quantum Inf. Comput. {\bf 7}, 459 (2007).

\bibitem{Wootters} W. K. Wootters, Phys. Rev. Lett. {\bf 80}, 2245 (1998).

\bibitem{Adesso} D. Girolami and G. Adesso, Phys. Rev. A {\bf 83}, 052108 (2011).

\bibitem{Zhou} X. Wu and T. Zhou, e-print arXiv: 1504.00129.

\bibitem{Chen} Qing Chen, Chengje Zhang, Sixia Yu, X.X. Yi and C.H. Oh, Phys. Rev. A {\bf 84}, 042313 (2011).

\bibitem{Huang} Y. Huang,  Rev. A {\bf 88}, 014302 (2013).

\bibitem{Namkug} M. Namkung, J. Chang, J. Shin, Y. Kwon,  e-print arXiv: 1404.6329. To appear in Int. J. Theor. Phys. 

\bibitem{Ali} M. Ali, A. R. P. Rau, and G. Alber, Phys. Rev. A {\bf 81}, 042105 (2010).

\bibitem{Luo} S. Luo, Phys. Rev. A {\bf 77}, 042303 (2008).

\bibitem{Li} B. Li, Z.-X. Wang, and S.-M. Fei Phys. Rev. A {\bf 83}, 022321 (2011).



\bibitem{He2011} Q.-L. He, J.-B. Xu, D.-X. Yao, and Y.-Q. Zhang, Phys Rev A \textbf{84}, 022312 (2011)

\end{thebibliography}
\end{document}